\documentclass[12pt]{article}
\usepackage{amsmath}
\usepackage{amscd}
\usepackage{amssymb}
\usepackage{array}

\newtheorem{definition}{Definition}[section]

\newtheorem{example}[definition]{Example}

\begin{document}

\rightline{CERN-PH-TH/2005-049}


\rightline{FTUV-05/0325}

\newcommand{\ka}{K\"ahler }

\newcommand{\R}{\mathbb{R}}
\newcommand{\C}{\mathbb{C}}
\newcommand{\Z}{\mathbb{Z}}
\newcommand{\Hb}{\mathbb{H}}

\newcommand{\rSU}{\mathrm{SU}}
\newcommand{\rU}{\mathrm{U}}
\newcommand{\rSO}{\mathrm{SO}}

\newcommand{\rspan}{\mathrm{span}}
\newcommand{\rsolv}{\mathrm{solv}}
\newcommand{\rtr}{\mathrm{tr}}

\newcommand{\fso}{\mathfrak{so}}
\newcommand{\fsu}{\mathfrak{su}}
\newcommand{\fg}{\mathfrak{g}}
\newcommand{\fh}{\mathfrak{h}}
\newcommand{\fp}{\mathfrak{p}}
\newcommand{\fk}{\mathfrak{k}}
\newcommand{\fs}{\mathfrak{s}}
\newcommand{\fid}{\mathfrak{i}}
\newcommand{\ft}{\mathfrak{t}}
\newcommand{\fn}{\mathfrak{n}}
\newcommand{\fa}{\mathfrak{a}}
\newcommand{\fsolv}{\mathfrak{solv}}

\newcommand{\cM}{\mathcal{M}}
\newcommand{\cI}{\mathcal{I}}
\newcommand{\cL}{\mathcal{L}}
\newcommand{\cK}{\mathcal{K}}
\newcommand{\cF}{\mathcal{F}}
\newcommand{\cN}{\mathcal{N}}
\newcommand{\cH}{\mathcal{H}}

\newcommand{\e}{\epsilon}

\newcommand{\id}{\relax{\rm 1\kern-.35em 1}}

\vskip 1cm

  \centerline{\LARGE \bf  Integration of massive states as}

  \bigskip

  \centerline{\LARGE \bf  contractions of non linear $\sigma$-models.}

\vskip 1.5cm

\centerline{L. Andrianopoli$^{\flat}$,  S. Ferrara$^{\sharp}$, M. A.
Lled\'o$^{\natural}$ and O. Maci\'a$^{\natural}$}
 \vskip
1.5cm

\centerline{\it $^\flat$ Centro E. Fermi, Compendio Viminale,
I-00184 Rome, Italy } \centerline{{\footnotesize e-mail:
Laura.Andrianopoli@cern.ch}}

 \medskip

 \centerline{\it $^\sharp$ Department
of Physics, Theory Division} \centerline{\it
 CERN, CH 1211 Geneva 23, Switzerland and }

\smallskip

\centerline{\it \it INFN, Laboratori Nazionali di Frascati, Italy.}
\centerline{{\footnotesize e-mail: Sergio.Ferrara@cern.ch}}
\medskip
\centerline{\it $^\natural$
 Departament de F\'{\i}sica Te\`orica,
Universitat de Val\`encia and IFIC}
 \centerline{\small\it C/Dr.
Moliner, 50, E-46100 Burjassot (Val\`encia), Spain.}
 \centerline{{\footnotesize e-mail: Maria.Lledo@ific.uv.es,\; Oscar.Macia@ific.uv.es}}


\vskip 2cm

\begin{abstract}
We consider the contraction of some non linear $\sigma$-models which
appear in effective supergravity theories. In particular we consider
the contractions of maximally symmetric spaces corresponding to
$N=1$ and $N=2$ theories, as they appear in certain low energy
effective supergravity actions with mass deformations.

The contraction procedure is shown to describe the {\em integrating
out} of massive modes in the presence of interactions, as it happens
in many supergravity models after spontaneous supersymmetry
breaking.
\end{abstract}

 \vfill\eject

\section{Introduction}

Supergravity theories with mass deformations have recently received
some attention  because of their relation to flux compactifications (for a review see, {\it e.g.}, \cite{schulz})
or Scherk--Schwarz generalized dimensional reduction \cite{ss}.

For $N\geq 2$ local supersymmetry, the supergravity theories admit
mass deformations that always  correspond to gauged supergravities \cite{fgp,cgp}.
The mass parameters may be chosen in such a way that a low energy
effective Lagrangian for the massless sector can be  singled out by
deleting the massive modes.

This procedure is usually discussed in the framework of consistent
truncations of field theories \cite{duff}, but here we want to show that the
same phenomenon may arise as well as a {\em contraction}. The basic
argument is that the limiting situation of a mass scale asymptotically
large is equivalent to the contraction of  some group structure.

Suppose the group structure is a non-linear $\sigma$-model related
to a maximally symmetric space $G/H$ where $G$ is non compact and
$H$ its maximally compact subgroup \cite{he,va}. One can make an
In\"{o}n\"u-Wigner contraction \cite{iw} of the group $G$ with
respect to a subgroup $G'$. Let $H'=H\cap G'$. We can induce a
contraction $G/H$ to a manifold which will have $G'/H'$ embedded
in it. The contracted manifold has the same dimension as the
original one (as it happens for contractions of algebras and
groups), but with a  metric that will be essentially different. An
example of this are the contractions of the hyperboloid
$\rSU(1,1)/\rU(1)$. If the contraction is made with respect to the
subgroup $\rU(1)$ one obtains the flat metric, while if the
contraction is made with respect to $\rSO(1,1)$ one obtains an
hyperbolic sheet, with one translational isometry. We will see in
detail how to compute the metrics in these and other cases.

 There are other types of contractions that do not fit in the scheme
 described above, but that may have physical interest. If $G/H$ is a symmetric
 space of the non compact type, it inherits a group structure through the Iwasawa
 decomposition of $G$
$$G= G_S\times H .$$ Then
 $G/H\approx G_S$ is the solvable Lie group associated to $G$ \cite{he,va,al}. Note
that $G_S$ depends on the real (non compact) form of $G$. We can
then consider  contractions in which $G/H$ goes to $G'/H' \ltimes
\R^n$ with $\dim(G/H) = \dim (G'/H') + n$, independently of the fact
that $G'$ is a subgroup of $G$ or not.  The physical interpretation of these contractions is as a (super)-Higgs mechanism \cite{adfl0}, where  the
massive modes are described by  $\R^n$ degrees of freedom while the fields which remain  massless  are
in $G'/H'$ \cite{fp}. Indeed, because of the semidirect product structure, it is
always consistent to set to zero (which, in this case, would correspond to integrate out) the elements of $\R^n$, since
$\R^n$ is an invariant subalgebra of $G'/H' \ltimes \R^n$ \cite{lou}. We will
consider several examples and discuss their physical applications.

The paper is organized as follows:

In Section 2 we describe the solvable algebras related to symmetric
spaces
  $$\frac{\rSO(1,1+n)}{\rSO(1+n)}, \quad
  \frac{\rSU(1,1+n)}{\rU(1) \times \rSU(1+n)},\quad \frac{\rSO(2,2+n)}{\rSO(2)\times \rSO(2+n)},$$
  $$\frac{\rSU(2,2+n)}{\rU(1)\times\rSU(2)\times \rSU(2+n)},$$ and how
 these algebras are embedded one into the
other. We also compute the metric of these spaces in the solvable
parametrization. We show a couple of examples where these spaces
are related to one another by gauging some isometries in the
corresponding supergravity models followed by an integration of
the massive modes.

In Section 3 we study some contractions of the solvable algebras
introduced and we show how they are related among themselves. We
compute the contracted metric by first giving a {\em deformation}
of it in terms of a parameter $\epsilon$. The limits
$\e\rightarrow 1$ and $\e\rightarrow 0$ correspond to the original
and the contracted spaces respectively. For an arbitrary $\e\neq
0$, the groups are isomorphic but we will see that it is not
possible, in general, to reabsorb the parameter into a
redefinition of the coordinates of the coset space. This means
that the spaces at $\e\neq 0$ are not isometric. We will show this
phenomenon in detail. We will see how it is possible to interpret
the gauging and integrating procedure of the examples treated in
Section 2 as a contraction followed by a quotienting by a
submanifold.

In Section 4 we describe the super Higgs phenomenon associated to
an effective N=2 supergravity theory with scalar manifold
$$\frac{\rSU(1,1+n)}{(\rU(1)\times\rSU(1+n))}\times\frac{\rSU(2,2+n)}{(\rU(1)\times\rSU(2)
\times \rSU(2+n)},$$ relating it to the contraction procedure
described in previous sections.

In the Appendix  we explain in more detail the parametrization
chosen to study these sigma models.


\section{Symmetric spaces, solvable parametrizations and
embeddings\label{symmetric}}

We first illustrate the calculation of the solvable Lie algebra
associated to a symmetric space of the non compact type with the
simple example in our list. Essentially one has to diagonalize
simultaneously the elements of the maximal abelian subalgebra in the
space $\fp$ of the Cartan decomposition
$$\fg=\fh+\fp, \quad \fg=\mathrm{Lie}(G),\quad
\fh=\mathrm{Lie}(H).$$

\subsection{Solvable parametrization of
$\rSO(1,1+n)/\rSO(1+n)$.\\}

We consider the Lie algebra of $\rSO(1,1+n)$, $\fso(1,1+n)$. In the
fundamental representation, an element of it is given by
$$X=\begin{pmatrix}A&\begin{matrix}b_1\\\vdots\\b_{n+1}\end{matrix}\\b_1\cdots
b_{n+1}&0\end{pmatrix}, \qquad A=-A^T$$  $A$ is an antisymmetric
$(n+1)\times(n+1)$ matrix. The Cartan decomposition of
$\fg=\fso(1,1+n)$ is
$$\fg=\fk+\fp, \qquad
\fk=\fso(1+n)=\left\{\begin{pmatrix}A&0\\0&0\end{pmatrix}\right\},
\quad \fp=\left\{\begin{pmatrix}0&b\\b^T&0\end{pmatrix}\right\}.$$
It is easy to see that the coset has rank one. We choose the
element
$$H=\begin{pmatrix}0&\begin{matrix}0\\\vdots\\0\end{matrix}&\begin{matrix}0\\\vdots\\0\end{matrix}
\\0\cdots
0&0&1
\\0\cdots
0&1&0\end{pmatrix}$$ as the generator of the maximal abelian
subalgebra in $\fp$. We must diagonalize $h$ to obtain the reduced
root pattern. This is easier by noting the following decomposition
$$\fso(1,n+1)\rightarrow
\fso(1,1)+\fso(n)+\mathbf{n}^++\mathbf{n}^-,$$ where
$$\fso(1,1)=\rspan\{H\}, \quad \mathbf{n}^\pm=\left\{
\begin{pmatrix}&0&&\begin{matrix}b_1\\\vdots\\b_n\end{matrix}&\begin{matrix}
 \mp b_1\\\vdots\\\mp
 b_n\end{matrix}\\-b_1&\cdots&-b_n&0&0\\\mp b_1&\cdots&\mp
 b_n&0&0\end{pmatrix}\right\},$$
 and a vector in $ \mathbf{n}^\pm$ has charge $\pm 1$ with
 respect to $H$. In this decomposition the algebra shows a
 $\fso(1,1)$ grading ($\fso(n)$ has degree 0), and $
 \mathbf{n}^\pm$ are nilpotent (in particular, abelian)
 subalgebras. The solvable Lie algebra associated to the coset
 $\rSO(1,1+n)/\rSO(1+n)$ is then
 \begin{equation}
 \fsolv\left(\frac{\rSO(1,1+n)}{\rSO(1+n)}\right)=\rspan\{H\}\ltimes
 \mathbf{n}^+,
 \label{so1n}
 \end{equation}
 with commutation rules
 $$[H,X_i]=X_i,\quad  i=1,\dots n\qquad \hbox{(the rest
 zero)}.$$
 Finally, the Iwasawa decomposition of the Lie algebra is
 $$\fso(1,1+n)=\fso(1+n)+\fsolv\left(\frac{\rSO(1,1+n)}{\rSO(1+n)}\right).$$

 \bigskip
 We choose a coset representative of the following form
 $$L=e^{u_iX_i}e^{\varphi H}, \qquad L^{-1}=e^{-\varphi
 H}e^{-u_iX_i}.$$ We will see that this kind of splitting of the generators is specially useful.
 The pull back of the Maurer-Cartan form on the
 group to the coset space, $L^{-1}dL$, decomposes as
 $$L^{-1}dL=(L^{-1}dL)_\fk+(L^{-1}dL)_\fp.$$ The first term is the
 connection on the $K$-bundle $G\rightarrow G/K$, with $K=\rSO(1+n)$ (spin
 bundle and spin connection) and the second term is the vielbein of $G/K$.

 The metric is then  computed as
 $$ds^2= \langle(L^{-1}dL)_\fp,(L^{-1}dL)_\fp\rangle$$
 where $\langle\;,\;\rangle$ is the Cartan-Killing form on $\fg$. Using the relation
 $$e^{\alpha X}Ye^{-\alpha X}=Ye^{\alpha \beta}, \quad \hbox{provided}
 \quad
 [X,Y]=\beta Y,$$
  it is easy to see that the metric becomes
 \begin{equation}ds^2=d\varphi^2+e^{-2\varphi}\sum_i{du_i}^2.\label{metriceasy}\end{equation}
 This metric has the translational isometries $u_i\rightarrow
 u_i+c_i$ which  are a maximal abelian ideal
 of the solvable Lie algebra (see  Appendix \ref{translational}).
 In this case the ideal is $\cI=\rspan\{X_i\}$.

\bigskip

It is now easy to go to the largest space in our list. It has only
rank two and the rest of the solvable algebras can be seen as
subalgebras of this. In fact we have a chain of embeddings of the
solvable Lie algebras which implies a chain of embeddings of the
corresponding symmetric spaces.


\subsection{Solvable parametrization of
${\rU(2,2+n)}/{\rU(2)\times \rU(2+n)}$.}

An element of the Lie algebra $\fsu(2,2+n)$ can be written as
 $$X=\begin{pmatrix}A_{2\times 2}&B_{2\times (2+n)}\\B^\dagger_{(2+n)\times 2}&
 D_{(2+n)\times(2+n)}\end{pmatrix}, \qquad
 A^\dagger=-A, \quad C^\dagger=\-C,$$
 and the Cartan decomposition of $\fsu(2,2+n)=\fh+\fp$ is
 $$\fh=\left\{\begin{pmatrix}A_{2\times 2}&0\\0&
 D_{(2+n)\times(2+n)}\end{pmatrix}\right\},\qquad \fp=\left\{
 \begin{pmatrix}0&B_{2\times (2+n)}\\B^\dagger_{(2+n)\times 2}&
 0\end{pmatrix}\right\}.$$
 A maximal abelian subalgebra of $\fp$ has dimension 2, and so the coset
 has rank two. We can choose for example, as maximal abelian
 subalgebra, the one generated by the matrices
$$H_+=\begin{pmatrix}0&0&0&1&0&\cdots&0\\
0&0&0&0&0&\cdots&0\\
0&0&0&0&0&\cdots&0\\
1&0&0&0&0&\cdots&0\\
\vdots&\cdots &\vdots\\
0&0&0&0&0&\cdots&0\\
\end{pmatrix}, \qquad H_-=\begin{pmatrix}0&0&0&0&0&\cdots&0\\
0&0&i&0&0&\cdots&0\\
0&-i&0&0&0&\cdots&0\\
0&0&0&0&0&\cdots&0\\
\vdots&\cdots &\vdots\\
0&0&0&0&0&\cdots&0\\
\end{pmatrix}$$

The solvable algebra can be shown to be generated by
\begin{eqnarray}\fs_4&=&\fsolv\left(\frac{\rU(2,2+n)}{\rU(2)\times\rU(2+n)}\right)=\rspan\{H_+,
H_-\}+\nonumber\\\nonumber \\&&\rspan\{ Z^{ia},
Y^{ia}, T^{2,0}, T^{0,2}, S_\alpha^{(1,1)},S_\alpha^{(1,-1)}\}, \label{s4}\\
\nonumber\\ &&\hbox{where}\quad  i=1,2,\quad  a=1,\dots n,\quad
\alpha=1,2,\nonumber\end{eqnarray}  with commutation rules
\begin{eqnarray}
&&[Z^{ia},Z^{jb}]=\epsilon^{ij}\delta^{ab}T^{(2,0)}\nonumber\\
&&[Y^{ia},Y^{jb}]=\epsilon^{ij}\delta^{ab}T^{(0,2)}\nonumber\\&&
[Z^{ia},Y^{jb}]=\delta^{ab}(\delta^{ij}S_2^{(1,1)}+\epsilon^{ij}S_1^{(1,1)})\nonumber\\
&&[Y^{ia}, S_1^{(1,-1)}]=Z^{ia}\nonumber\\
&&[Y^{ia}, S_2^{(1,-1)}]=\epsilon^{ij}Z^{ja}\nonumber\\
&&[T^{(0,2)}, S_\alpha^{(1,-1)}]=2S_\alpha^{(1,1)}\nonumber\\
&&[S_\alpha^{(1,1)},S_\beta^{(1,-1)}]=\delta_{\alpha\beta}T^{(2,0)}\nonumber\\
&&[H_+, Z^{ia}]=Z^{ia}\nonumber\\
&&[H_-, Y^{ia}]=Y^{ia}.\label{su23}
\end{eqnarray}
The rest of the commutators with the Cartan generators $H_+$ and
$H_-$ are indicated by the superindices $(h_+,h_-)$. All the other
commutators are zero.

%

Based on this solvable algebra, we choose the following
parametrization for the coset representative of
   ${\rSU(2,2+n)}/({\rSU(2)\times\rSU(2+n)\times\rU(1)})$ \footnote{In the appendix we show that
    the generators $T^{(2,0)}, T^{(0,2)}, S^{(1,1)}_\alpha, Z^{1a}$ correspond to true global
     {\em translational isometries}.}:
   \begin{equation}
   L(t,\tilde t, \tilde s_\alpha, s_\alpha, z_{ia}, y_{ia},
   \psi,\phi)=A(t,\tilde t, \tilde s_\alpha,
   z_{1a})B(s_\alpha, z_{2a},
   y_{ia})C(\psi,\phi)\label{solvpar}\end{equation} where
   \begin{eqnarray*}A&=&\exp{(tT^{(2,0)}+\tilde t T^{(0,2)}+\tilde
   s_\alpha S^{(1,1)}_\alpha+z_{1a} Z^{1a})}\\
   B&=& \exp{(s_1S_1^{(1,-1)})}\exp{(s_2S_2^{(1,-1)})}
   \exp({z_{2a}Z^{2a}})\exp({y_{2a}Y^{2a}})\exp({y_{1a}Y^{1a}})\\
   C&=& \exp{( \psi H_+ + \phi H_-)}
   \end{eqnarray*}

   The Maurer Cartan form is

   \begin{eqnarray}
   L^{-1}dL&=&e^{-2\psi}(s_\alpha d\tilde s_\alpha  + (s_1^2+s_2^2)
   d\tilde t +dt +z_{2a}
   dz_{1a})T^{(2,0)}+e^{-2\phi}(d\tilde t
   -y_{1a}dy_{2a})T^{(0,2)}+\nonumber\\&&+ e^{-(\phi+\psi)}(y_{1a}y_{2a}ds_1+\frac12 (y_1^2+y_2^2)ds_2 +
   d\tilde s_1 + 2s_1d\tilde t
   -\epsilon_{ij}y_{ia}dz_{ja})S^{(1,1)}_1+\nonumber\\&&+ e^{-(\phi +\psi)}(y_{1a}y_{2a}ds_2-\frac12
   (y_1^2+y_2^2)ds_1+d\tilde s_2 + 2 s_2 d\tilde t+
   \delta_{ij} y_{ia}dz_{ja})S^{(1,1)}_2+\nonumber\\&&
   + e^{-\psi}
   (-y_{1a}ds_1+y_{2a}ds_2+dz_{1a})Z^{1a}+e^{-\psi}(-y_{2a}ds_1-y_{1a}ds_2+dz_{2a})Z^{2a}+\nonumber\\&&
   +e^{\phi-\psi}ds_\alpha S^{(1,-1)}_\alpha +
   e^{-\phi}dy_{ia}Y^{ia}+d\psi H_+ + d\phi H_-\label{vielbein}
   \end{eqnarray}

The metric of the coset is  computed now as
   $\langle(L^{-1}dL)_\fp, (L^{-1}dL\rangle_\fp)$. On the tangent
   space to the identity, this gives the following inner product:
   \begin{eqnarray*}
   \langle X,X\rangle=1, &\quad & \hbox{for } X=H_\pm, T^{(2,0)},T^{(0,2)}\\
 \langle X,X\rangle=\frac 12,&\quad & \hbox{for }
 X=S_1^{(1,1)},S_2^{(1,1)},S_1^{(1,-1)},S_2^{(1,-1)}, Y^{1a},
 Y^{2a},Z^{1a}, Z^{2a},
   \end{eqnarray*}
and the rest zero.
   \bigskip

   For $n=0$ we
   obtain the reduced expression

 \begin{eqnarray}
   &&ds^2=
   d\phi^2 +
   d\psi^2+
   e^{-4\psi}dtdt+
   2e^{-4\psi} s_1 dtd\tilde s_1 +
   2 e^{-4\psi} s_2 dtd\tilde s_2\nonumber\\
   &&  +
   2e^{-4\psi}(s_2^2 + s_1^2)dtd\tilde t+
   \frac{1}{2}(e^{-2(\psi+\phi)}+ 2e^{-4\psi}s_1^2)d\tilde s_1 d\tilde s_1
   \nonumber\\
   && +2 e^{-4\psi} s_2  s_1 d\tilde s_1d\tilde s_2
    + 2 e^{-4\psi} s_1 (e^{2(\psi-\phi)}+ s_2^2 + s_1^2 ) d\tilde
   s_1d\tilde t \nonumber\\
   &&+
 \frac{1}{2}(e^{-2(\psi+\phi)}+2e^{-4\psi} s_2^2)d\tilde s_2 d\tilde
   s_2 \nonumber \\&&+ 2e^{-4\psi}s_2( e^{2(\psi-\phi)}+ s_2^2 +  s_1^2 )d\tilde s_2
   d\tilde t\nonumber\\
   && +
   \frac{1}{2}e^{2(\phi-\psi)}d s_2 d s_2 + \frac{1}{2}e^{2(\phi-\psi)}d s_1 d s_1\nonumber\\
   &&+ e^{-4(\psi+\phi)}\left(e^{2\psi}+e^{2\phi}( s_1^2 + s_2^2)
   \right)^2 d\tilde t d\tilde t.\label{bigmetricn=0e=1}
   \end{eqnarray}

For arbitrary $n$ we obtain \footnote{For this and the rest of the
calculations of different metrics we have used the program Wolfram
Research, Inc., Mathematica, Version 5.1, Champaign, IL (2004).}
(sum over repeated indices is understood, and we have used the
short-hand notation $y_1^2=y_{1a}y_{1a}$):

\begin{eqnarray}
   &&ds^2=
   d\phi^2 +
   d\psi^2+
   e^{-4\psi}dtdt+
   2e^{-4\psi} s_1 dtd\tilde s_1 +
   2 e^{-4\psi} s_2 dtd\tilde s_2\nonumber\\
   && +2e^{-4\psi}z_{2a} dtdz_{1a} +
   2e^{-4\psi}(s_2^2 + s_1^2)dtd\tilde t+
   \frac{1}{2}(e^{-2(\psi+\phi)}+ 2e^{-4\psi}s_1^2)d\tilde s_1 d\tilde s_1
   \nonumber\\
   && +2 e^{-4\psi} s_2  s_1 d\tilde s_1d\tilde s_2  +
   \frac{1}{2} e^{-2(\psi+\phi)}(y_{1}^2+y_{2}^2)d\tilde s_1d s_2+
   e^{-2(\psi+\phi)}y_{1a}y_{2a} d\tilde s_1d s_1 \nonumber\\
   && -e^{-2(\psi+\phi)}y_{1a} d\tilde s_1 dz_{2a}+
   (2e^{-4\psi} s_1 z_{2a} +e^{-2(\psi +\phi)}y_{2a} ) d\tilde
   s_1dz_{1a}\nonumber\\
   && + 2 e^{-4\psi} s_1 (e^{2(\psi-\phi)}+ s_2^2 + s_1^2 ) d\tilde
   s_1d\tilde t +
 \frac{1}{2}(e^{-2(\psi+\phi)}+2e^{-4\psi} s_2^2)d\tilde s_2 d\tilde
   s_2\nonumber\\
   &&+e^{-2(\psi+\phi)}y_{1a}y_{2a} d\tilde s_2 d s_2
   -\frac{1}{2}e^{-2(\psi+\phi)}(y_{1}^2 +y_{2}^2)d\tilde s_2 d s_1\nonumber\\
   &&+e^{-2(\psi+\phi)}y_{2a}d\tilde s_2dz_{2a}+
   (2e^{-4\psi}s_2 z_{2a}+e^{-2(\psi+\phi)}y_{1a})d\tilde s_2dz_{1a} \nonumber\\
   && + 2e^{-4\psi}s_2( e^{2(\psi-\phi)}+ s_2^2 +  s_1^2 )d\tilde s_2
   d\tilde t\nonumber\\
   && +
   \frac{1}{8}e^{-2(\psi+\phi)}(4e^{4\phi}+ 4e^{2\phi}
   (y_{1}^2+y_{2}^2)+ 4(y_{1a}y_{2a})^2+(y_{1}^2+y_{2}^2)(y_{1}^2+y_{2}^2) )d s_\alpha d s_\alpha \nonumber\\
   && - \frac{1}{2}e^{-2(\psi+\phi)}(2e^{2\phi}y_{1b}+(-2(y_{1a}y_{2a})y_{2b}+(y_{1}^2
   +y_{2}^2)y_{1b}))d s_2
   dz_{2b}\nonumber\\&&+
   \frac{1}{2}e^{-2(\psi+\phi)}(2e^{2\phi}y_{2b}+(2(y_{1a}y_{2a})y_{1b}+
   (y_{1}^2+y_{2}^2)y_{2b}))d s_2
   dz_{1b}\nonumber\\
   && + e^{-2(\psi+\phi)}(y_{2}^2 s_1+2y_{2a}  s_2
   y_{1a}+ s_1 y_{1}^2)d s_2 d\tilde t\nonumber\\
    && - \frac{1}{2}e^{-2(\psi+\phi)}(2e^{2\phi}y_{2b}+(2(y_{1a}y_{2a})y_{1b}+(y_{1}^2+y_{2}^2)y_{2b}))
    d s_1dz_{2b}\nonumber \\&&
   - \frac{1}{2}e^{-2(\psi+\phi)}(2e^{2\phi}y_{1b}+(-2(y_{1a}y_{2a})y_{2b}+
   (y_{1}^2+y_{2}^2)y_{1b}))d s_1 dz_{1b}
   \nonumber\\
   &&-e^{-2(\psi+\phi)}(y_{1}^2  s_2 - 2y_{1a} s_1 y_{2a} +  s_2 y_{2}^2 )
   ds_1d\tilde t\nonumber\\&&-
\frac{1}{2}e^{-2(\phi+\psi)}\epsilon_{ij}\epsilon_{mn}(y_{ia}y_{jb})dz_{ma}dz_{nb}\nonumber\\&&
   +\frac{1}{2}e^{-2(\psi+\phi)}(e^{2\phi}\delta_{ab}+(y_{1a}y_{1b}+y_{2a}y_{2b}))dz_{2a}dz_{2b}\nonumber\\
 && -e^{-2(\psi+\phi)}(2y_1 s_1 - 2  s_2 y_2) dz_2
   d\tilde t\nonumber\\&&+
   \frac{1}{2}e^{-4\psi}(e^{2\psi}\delta_{ab}+2z_{2a}z_{2b}+e^{2(\psi-\phi)}
   (y_{1a}y_{1b}+y_{2a}y_{2b}))dz_{1a} dz_{1b}\nonumber\\
   && +(2e^{-4\psi}( s_2^2+ s_1^2)z_{2a} +2e^{-2(\psi+\phi)}(y_{1a} s_2+ s_1
   y_{2a}))dz_{1a}d\tilde t\nonumber\\
   &&+ e^{-4(\psi+\phi)}(e^{2\psi}+e^{2\phi}( s_1^2 + s_2^2)
   )^2 d\tilde t d\tilde t- 2 e^{-4\phi}y_{1a} d\tilde tdy_{2a}\nonumber\\
   && + \frac{1}{2}e^{-4\phi}(e^{2\phi}\delta_{ab}+2y_{1a}y_{1b})dy_{2a} dy_{2b} +
   \frac{1}{2}e^{-2\phi}dy_{1a}dy_{1a} \label{bigmetrice=1}
   \end{eqnarray}

\subsection{Chain of embeddings}

We  have the following chain of solvable Lie algebras

\begin{eqnarray}
\fs_4&=&\fsolv\left(\frac{\rSU(2,2+n)}{\rU(1)\times\rSU(2)\times\rSU(2+n)}\right)\qquad
\hbox{(see (\ref{s4}))},\nonumber\\
\fs_3&=&\fsolv\left(\frac{\rSO(2,2+n)}{\rSO(2)\times\rSO(2+n)}\right)=\rspan\{H_+,
H_-\}+ \rspan\{ Z^{1a}, Y^{1a}, S_2^{(1,1)},S_2^{(1,-1)}\},\nonumber\\
\fs_2&=&\fsolv\left(\frac{\rSU(1,1+n)}{\rU(1)\times\rSU(1+n)}\right)=\rspan\{H_++
H_-\}+\rspan\{ Z^{1a}, Y^{1a}, S_2^{(1,1)}\},\nonumber\\
\fs_1&=&\fsolv\left(\frac{\rSO(1,1+n)}{\rSO(1+n)}\right)=\rspan\{H_++
H_-\}+\rspan\{ Y^{1a}\},\label{chainsolvable}
\end{eqnarray}
with $\fs_i\subset_{\mathrm{sub}} \fs_{i+1}$. Following the same
procedure than in the previous examples, one can  show that these
solvable Lie algebras correspond to the following chain of symmetric
spaces:
 \begin{equation}\frac{\rSO(1,1+n)}{\rSO(1+n)}\subset\frac{\rSU(1,1+n)}{\rU(1+n)}
 \subset \frac{\rSO(2,2+n)}{\rSO(2)\times\rSO(2+n)}
 \subset \frac{\rSU(2,2+n)}{\rU(2)\times\rSU(2+n)}\label{embeddings}\end{equation}

 Notice that in this chain we have
 $$\frac {G_i}{H_i}\subset \frac {G_{i+1}}{H_{i+1}}$$ with
 \begin{eqnarray*}
 G_1\subset G_{2}, \quad &&\quad  H_1\subset H_{2},\\
 G_3\subset G_{4}, \quad &&\quad  H_3\subset H_{4},\\
 G_1\subset G_{3}, \quad &&\quad  H_1\subset H_{3},\\
 G_2\subset G_{4}, \quad &&\quad  H_2\subset H_{4},\\
\\\end{eqnarray*} but $G_2$ is not in $G_{3}$ nor $H_2$ in
$H_{3}$ for generic $n$.

\bigskip

The solvable parametrization  (\ref{solvpar}) allows us to compute
the metric of the spaces in (\ref{embeddings}) by imposing
different restrictions on (\ref{bigmetrice=1}).

For the coset $\rSO(2,2+n)/(\rSO(2)\times\rSO(2+n))$, we have
$$z_{2a}=y_{2a}=t=\tilde t=\tilde s_1=s_2=0,$$ so the metric is

\begin{eqnarray}
ds^2&=& d\phi^2 + d\psi^2 +\frac{1}{2}e^{-2(\psi+\phi )}d\tilde
s_2 d\tilde s_2 -\frac{1}{2}e^{-2(\psi+\phi)}y_1^2 d\tilde s_2 d
s_1\nonumber\\
&& +e^{-2(\psi+\phi)}y_{1a}d\tilde s_2dz_{1a}
 + \frac{1}{8}e^{-2(\psi+\phi)}(4e^{4\phi}+(y_1^2)^2+4e^{2\phi}y_1^2)d
 s_1 d s_1
 \nonumber\\&&- \frac{1}{2}e^{-2(\psi+\phi)}y_{1a}(2e^{2\phi}+y_{1}^2
)d s_1 dz_{1a} +\frac{1}{2}e^{-4\psi}(e^{2\psi}\delta_{ab}
+e^{2(\psi-\phi)}y_{1a}y_{1b})dz_{1a} dz_{1b}\nonumber\\
&&+ \frac{1}{2}e^{-2\phi}dy_{1a}dy_{1a}. \label{metric3}
\end{eqnarray}
Imposing the constraints $s_1=\phi-\psi=0$, we obtain the metric on
$\rSU(1,1+n)/\rU(1+n)$,
\begin{eqnarray}
ds^2&=& 2d\phi^2 +\frac{1}{2}e^{-4\phi}d\tilde s_2 d\tilde s_2 +
e^{-4\phi}y_{1a} d\tilde s_2dz_{1a}
 \nonumber\\&&
+\frac{1}{2}e^{-4\phi}(e^{2\phi} \delta_{ab} +y_{1a}y_{1b})dz_{1a}
dz_{1b} +\frac{1}{2}e^{-2\phi}dy_{1a}dy_{1a},\label{metric2}
\end{eqnarray}
and imposing $z_{1a}=\tilde s_2=0$, we obtain the metric for
$\rSO(1,1+n)/\rSO(1+n)$:
\begin{eqnarray}
ds^2= 2d\phi^2+\frac{1}{2}e^{-2\phi}dy_{1a}
dy_{1a}\label{metric1}\end{eqnarray} which, up to a rescaling of
the coordinates, corresponds to (\ref{metriceasy}).

We can further impose $y_{1n}=0$ to obtain the same form than
(\ref{metric1}) but with $a=1,\dots n-1$. It is the metric of
$\rSO(1,n)/\rSO(n)$.

\subsection{Truncations and integration of massive modes \label{integration}}

Let us consider a sigma model described by the metric
(\ref{metric1}). As we have seen, this model has $n$ translational
isometries corresponding to the coordinates $y_{1a}$. We may
consider gauging one of these isometries, say $y_{1n}$. We
introduce a gauge field $A=A_\mu dx^\mu$ and substitute $dy_{1n}$
by the covariant differential
$$Dy_{1n}=dy_{1n}+gA.$$
We redefine the connection by a gauge transformation
$$\hat A=A+\frac 1gdy_{1n},$$
which will not change the kinetic term for $A$. Substituting this
definition in the metric we obtain
 $$ds^2=2d\phi^2+\frac{1}{2}e^{-2\phi}\sum_{a=1}^{n-1}dy_{1a}
dy_{1a}+\frac{1}{2}e^{-2\phi}g^2\hat A^2.$$ We see that the effect
of the gauging is absorbing the field $y_{1n}$ to give mass to the
gauge vector. Moreover, in this model $\hat A$ is decoupled from
the rest of fields (except for the warping factor $e^{-2\phi}$),
so setting $\hat A=0$ is consistent with the equations of motion.
After the truncation  the sigma model becomes $\rSO(1,n)/\rSO(n)$.
This is explained by the mathematical identity
$$\fsolv\left(\frac{\rSO(1,1+n)}{\rSO(1+n)} \right)= \fsolv\left(\frac{\rSO(1,1+n-k)}{\rSO(1+n-k)} \right)
\ltimes \R^k$$
which is a consequence of (\ref{so1n}).
\bigskip

We want to consider now the model $\rSU(1,2)/\rU(2)$, with metric
(\ref{metric2}) for $n=1$. Note that $\tilde s_2$ and $z_1$ are
translational isometries. As before, we can gauge them  by
introducing abelian connections $A^1$, $A^2$ with covariant
differentials
\begin{eqnarray*} d\tilde s_2&\rightarrow&D\tilde s_2= d\tilde
s_2+k_1A^1\\
dz_1&\rightarrow & Dz_1= dz_1+k_2A^2.\end{eqnarray*} We define the
gauge-transformed connections
\begin{eqnarray*} \hat A^1&=A^1+\frac 1{k_1}d\tilde s_2\\
 \hat A^2&=A^2+\frac 1{k_2}dz_1.\end{eqnarray*}

 By substituting this definition, we can see that in the metric there
  will appear the terms
 $$ds^2=\cdots +\frac{1}{2}e^{-4\phi}(k_1)^2(\hat A^1)^2+\cdots
 +\frac{1}{2}e^{-2\phi}(k_2)^2(\hat A^2)^2+\cdots$$
 So, as before, the effect of the gauging has been to give mass to the
 vectors by absorbing the modes associated to the translational
 isometries.

Nevertheless, in this case other interactions are present. By
assuming that the mass of the vectors is big enough we can take
their kinetic terms to zero, and then we obtain algebraic
equations for $\hat A^1$, $\hat A^2$. A straightforward
calculation shows that, after the elimination of these fields the
metric that remains is $\rSO(1,2)/\rSO(2)$, that is eq.
(\ref{metric1}) with $n=1$.

\bigskip

The difference between the two models here described is that in
the first case the integration of the massive modes is exact (that
is, it is a consistent truncation of the theory), while in the
second case a limiting process is involved (masses $\rightarrow
\infty$).

In the next section we will see that these integrations can be
modeled by a {\em contraction} of the metric of the initial
manifold, followed by a {\em quotienting} of the manifold by a
submanifold.

\section{Contractions of groups and coset spaces\label{contract}}

\paragraph{Contraction of a Lie algebra with respect to a
subalgebra.\\}

We describe the In\"on\"u-Wigner contraction of an algebra with
respect to a subalgebra. Let $\fg$ be an arbitrary, finite
dimensional Lie algebra with commutator $[\,,\,]$ and let
$\fg=\fg_1+\fg_2$, with $\fg_1$ a subalgebra. We define the
following family of linear maps
$$\begin{CD}\phi_\epsilon:\fg@>>>\fg\\
x=x_1\oplus x_2@>>> x=x_1\oplus \epsilon x_2,\end{CD}$$ labelled by
a real parameter $\epsilon$, In matrix form, the map and its inverse
($\epsilon\neq 0$) are block-diagonal
$$\phi_\epsilon= \begin{pmatrix}\id&0\\0&\epsilon\id\end{pmatrix}, \qquad
\phi^{-1}_\epsilon= \begin{pmatrix}\id&0\\0&\frac
1\epsilon\id\end{pmatrix}.$$  We can define a new commutator
$$[X,Y]_\epsilon=
\phi^{-1}_\epsilon([\phi_\epsilon(X),\phi_\epsilon(Y)]), \qquad
X,Y\in \fg.$$ $[\,,\,]_\epsilon$ is a {\em deformed bracket}, but
of a simple form, since for $\e\neq 0$ is,  by construction,
isomorphic to the  bracket with $\e=1$. We define the {\it
contraction} of $\fg$ with respect to the subalgebra $\fg_1$ as a
Lie algebra with the same supporting vector space
$\fg_c\approx\fg$ and with commutator
\begin{equation}[X,Y]_c=\lim_{\epsilon\rightarrow 0}
\phi^{-1}_\epsilon([\phi_\epsilon(X),\phi_\epsilon(Y)]), \qquad
X,Y\in \fg.\label{contractedbracket}\end{equation} This bracket is
well defined but, since $\phi_\epsilon$ is not invertible,
$[\,,\,]_c$ will not be, in general, isomorphic to the original
bracket.


\paragraph{Representations of the contracted algebra.\\}

We consider now a representation of $\fg$ on a finite dimensional
vector space $W$
$$R(X):W\rightarrow W,\qquad X\in\fg$$ and
assume that $W=W_1\oplus W_2$ with $W_1$ an invariant subspace
under the action of the subalgebra $\fg_1$. As before, we define a
one parameter family of linear maps
$$\begin{CD}\psi_\epsilon:W@>>>W\\
w=w_1\oplus w_2@>>> w=w_1\oplus \epsilon w_2,\end{CD}$$ so
$$\psi_\epsilon= \begin{pmatrix}\id&0\\0&\epsilon\id\end{pmatrix}, \qquad
\psi^{-1}_\epsilon= \begin{pmatrix}\id&0\\0&\frac
1\epsilon\id\end{pmatrix}.$$ Let us denote
$$R_\epsilon(X)=\psi^{-1}_\epsilon \circ
R(\phi_\epsilon(X))\circ\psi_\epsilon, \qquad X\in \fg.$$
$R_\epsilon$ is a representation of the deformed algebra. It is
easy to check that the map $R_c$
$$R_c(X)=\lim_{\epsilon\rightarrow 0}R_\epsilon(X)$$
is a representation of $\fg_c$ on $W$.

Notice that $\psi_\epsilon=\phi_\epsilon$ for the adjoint
representation.

\paragraph{Generalized contractions\\}

The map $\phi_\e$ can in fact be more general than the one
considered before, the only constraint being that the bracket in
(\ref{contractedbracket}) is well defined. The conditions for this
to happen were studied in Ref. \cite{ww} and are called {\em
generalized In\"on\"u-Wigner contractions}. They are also a
particular example of {\em algebra expansions} \cite{dipv}.

We will use particular examples of generalized contractions where the
brackets can be seen explicitly to have a well defined limit.
We will not describe the general theory of these contractions, for
which we refer to the original paper, Ref.  \cite{ww}.


\subsection{Deformations and contractions of the metric: some examples.}
As we have seen, we can always contract an algebra $\fg$ with
respect to a subalgebra $\fg'$. The contracted algebra, $\fg_c$ will
have always the structure of a semidirect product
$$\fg_c=\fg'\ltimes\R^n.$$
In the chain (\ref{chainsolvable}) we have described subalgebras
of $\fs_4$, so we can contract each algebra $\fs_{i}$ with respect
to $\fs_{j}$ with $j<i$

\bigskip
Since the solvable Lie algebras are related to the corresponding
symmetric spaces, we are going to define a procedure to contract
the symmetric spaces.  We will start with a representation $R_\e$
of the deformed Lie algebra, and  compute the coset representative
as in (\ref{solvpar}) with this new representation. From this, one
can compute a deformed vielbein and a deformed metric. This
procedure will introduce the parameter $\e$ in the metric, so we
will have a uniparametric family of metrics. Then, we can take the
limit $\e\rightarrow 0$.

\bigskip

We are interested in the simple examples presented in
\ref{integration}. The first one is trivial, since the contraction
of $\fs_1$ for arbitrary $n$ by the subalgebra $\fs_1$ for $n-1$
has no effect, giving again $\fs_1$ for $n$.

Let us see how this works with the next  example. We start with
the algebra $\fsolv(\rSU(1,2)/\rU(2))$ (which is $\fs_2$ for
$n=1$) and we will work out the contraction with respect to
$\fsolv(\rSO(1,2)/\rSO(2))$ which is ($\fs_1$ for $n=1$).
\begin{eqnarray}\fs_2(n=1)&=&\rspan\{H_1=H_++H_-\}\,+\,\rspan\{Z^1, Y^1,
S_1^{(1,1)}\} , \\
\fs_1(n=1)&=&\rspan\{H_1\}\,+\,\rspan\{Y^1\}\label{s1n=1}.\end{eqnarray}
 It is convenient to write explicitly the commutation
rules
$$[H_1,Z^1]=Z^1, \quad [H_1,Y^1]=Y^1,\quad [H_1, S_1^{(1,1)}]=2S_1^{(1,1)},\quad [Z^1,Y^1]=S_1^{(1,1)}.$$

The deformed algebra is
$$[H_1,Z^1]_\e=Z^1, \quad [H,Y^1]_\e=Y^1,\quad [H_1, S_1^{(1,1)}]_\e=2S_1^{(1,1)},
\quad [Z^1,Y^1]_\e=\e S_1^{(1,1)}\rightarrow 0.$$ The contracted
algebra has the property that the only elements in $\fg'$ that act
on the  abelian factor $\R^n$ are the elements of the commuting
subalgebra of $\fp$. In our case this subalgebra
 is $H_1=H_++H_-$. This property will translate in a particularly
 simple form of the metric.

We consider the three dimensional representation (induced from the
fundamental of $\fsu(1,2)$). We decompose the representation space
as
$$\C^3=V_1\oplus V_2,\qquad
V_1=\{\begin{pmatrix}v_1\\0\\v_3\end{pmatrix}\}, \quad
V_2=\{\begin{pmatrix}0\\v_2\\0\end{pmatrix}\},$$ being $V_1$ an
invariant subspace under the subalgebra (\ref{s1n=1})and consider
the linear map $\psi_\epsilon(e_1\oplus e_2)=e_1\oplus \epsilon
e_2$. Then we have a three dimensional representation of the
deformed algebra,
\begin{eqnarray*}
&&R_\epsilon(H_1)=\begin{pmatrix}0&0&1\\0&0&0\\1&0&0\end{pmatrix},\\
&&R_\epsilon(Z^1)=\begin{pmatrix}0&i&0\\i\epsilon^2&0&-i\epsilon^2\\0&i&0\end{pmatrix}\rightarrow
\begin{pmatrix}0&i&0\\0&0&0\\0&i&0\end{pmatrix},\\
&&R_\epsilon(Y^1)=\begin{pmatrix}0&1&0\\-\epsilon^2&0&\epsilon^2\\0&1&0\end{pmatrix}\rightarrow
\begin{pmatrix}0&1&0\\0&0&0\\0&1&0\end{pmatrix},\\
&&R_\epsilon(S_1^{(1,1)})=-\epsilon\begin{pmatrix}i&0&0\\0&i&0\\0&0&-2i\end{pmatrix}\rightarrow
0.
\end{eqnarray*}

We compute now the vielbein and the metric in the way that we
indicated in section \ref{symmetric}. Notice that the Euclidean
metric that we put on the solvable Lie algebra with the deformed
bracket is {\it the same} than the one for $\e=1$. In this way the
normal metric Lie algebras (in the terminology of Ref. \cite{al})
are not isomorphic, nor are isometric the corresponding Riemannian
spaces. We obtain then a true deformation of the metric.

The result is
\begin{eqnarray}
ds^2&=& 2d\phi^2 +\frac{1}{2}e^{-4\phi}d\tilde s_2 d\tilde s_2 +
\e^2e^{-4\phi}y_1 d\tilde s_2dz_1
 \nonumber\\&&
+\frac{1}{2}e^{-4\phi}(e^{2\phi} +\e^4y_1^2)dz_1 dz_1
+\frac{1}{2}e^{-2\phi}dy_1dy_1,\label{metricricci}
\end{eqnarray}
which can be compared with (\ref{metric2}) for $\e=1$.

For $\e\rightarrow 0$ we get
\begin{eqnarray}
ds^2&=& \left(2d\phi^2
+\frac{1}{2}e^{-2\phi}dy_1dy_1\right)+\frac{1}{2}e^{-2\phi}dz_1dz_1+\frac{1}{2}e^{-4\phi}d\tilde
s_2 d\tilde s_2 . \label{csm1}\end{eqnarray}
 The first two factors correspond to
(\ref{metric1}). The remaining modes appear decoupled except for
warping factors of type $e^{a\phi}/2$. Then, imposing the
constraints $z_1=0=\tilde s_2$ is always a consistent truncation
of the contracted sigma model (\ref{csm1}). We see with this
simple example that integrating out massive modes can be
geometrically modeled by a contraction of the sigma model,
followed by a quotienting by the decoupled modes.

It is instructive to compute Ricci tensor of the deformed metric
(\ref{metricricci}). We obtain (in the ordered basis
$\phi,y_1,z_1,\tilde s_2$)
\begin{eqnarray}
R^a_b(\epsilon)=\left(
\begin{array}{cccc}
-6&0&0&0\\
0&-2(2+\epsilon^4)&0&0\\
0&0&-2(2+\epsilon^4)&0\\
0&0&8 y_1 \epsilon^2(\epsilon^4-1)&2(-4+\epsilon^4)
\end{array}\right)
\end{eqnarray}
We see that for arbitrary $\epsilon$ it is not an Einstein space.
In the relevant limits
$$
R^a_b(1)=\begin{pmatrix}
-6&0&0&0\\
0&-6&0&0\\
0&0&-6&0\\
0&0&0&-6
\end{pmatrix}, \qquad R^a_b(0)=
\begin{pmatrix}-6&0&0&0\\
0&-4&0&0\\
0&0&-4&0\\
0&0&0&-8 \end{pmatrix}$$
 For $\e=1$ we have an Einstein space, but not for arbitrary $\e$. It becomes clear from this
 simple example that the deformation cannot be reabsorbed by a
 change of coordinates.

\vfill\eject

We consider now the In\"onu-Wigner contraction of $\fs_4$ with
respect to $\fs_3$. For simplicity, we take $n=1 $, so we have
$\fs_4=\fs_3 +\fg$ where
\begin{eqnarray*}\fs_3&=&\rspan\{H_+,H_-,S^{(1,1)}_2,S^{(1,-1)}_1,Y^1,Z^1\}\\
\fg&=&\rspan\{T^{(2,0)},T^{(0,2)},S^{(1,1)}_1,S^{(1,-1)}_2,Y^2,Z^2\}\end{eqnarray*}

 Differently from the first example, we
use the adjoint representation to introduce the parameter $\e$.
The result for the metric is

   \begin{eqnarray}
   &&ds^2=
   d\phi^2 +
   d\psi^2+
   e^{-4\psi}dtdt+
   2e^{-4\psi} s_1 dtd\tilde s_1 +
   2 e^{-4\psi} s_2 dtd\tilde s_2\nonumber\\
   && +2e^{-4\psi}z_2 dtdz_1 +
   2e^{-4\psi}(s_2^2 \epsilon^2 + s_1^2)dtd\tilde t+
   \frac{1}{2}(e^{-2(\psi+\phi)}+ 2e^{-4\psi}s_1^2)d\tilde s_1 d\tilde s_1
   \nonumber\\
   && +2 e^{-4\psi} s_2  s_1 d\tilde s_1d\tilde s_2  +
   \frac{1}{2} e^{-2(\psi+\phi)}(y_1^2+y_2^2\epsilon^2)d\tilde s_1d s_2+
   e^{-2(\psi+\phi)}y_1y_2 d\tilde s_1d s_1 \nonumber\\
   && -e^{-2(\psi+\phi)}y_1 d\tilde s_1 dz_2+
   (2e^{-4\psi} s_1 z_2 +e^{-2(\psi +\phi)}y_2 ) d\tilde
   s_1dz_1\nonumber\\
   && + 2 e^{-4\psi} s_1 (e^{2(\psi-\phi)}+ s_2^2 \epsilon^2+ s_1^2 ) d\tilde
   s_1d\tilde t +
 \frac{1}{2}(e^{-2(\psi+\phi)}+2e^{-4\psi} s_2^2)d\tilde s_2 d\tilde
   s_2\nonumber\\
   &&+e^{-2(\psi+\phi)}y_1y_2\epsilon^2 d\tilde s_2 d s_2
   -\frac{1}{2}e^{-2(\psi+\phi)}(y_1^2 +y_2^2\epsilon^2)d\tilde s_2 d s_1\nonumber\\
   &&+e^{-2(\psi+\phi)}y_2 \epsilon^2d\tilde s_2dz_2+
   (2e^{-4\psi}s_2 z_2+e^{-2(\psi+\phi)}y_1)d\tilde s_2dz_1 \nonumber\\
   && + 2e^{-4\psi}s_2( e^{2(\psi-\phi)}\epsilon^2+ s_2^2\epsilon^2 +  s_1^2 )d\tilde s_2
   d\tilde t\nonumber\\
   && +
   \frac{1}{8}e^{-2(\psi+\phi)}(4e^{4\phi}+y_1^4+(2\epsilon^2+4\epsilon^4)y_1^2y_2^2+y_2^4\epsilon^4+4e^{2\phi}
   (y_1^2+y_2^2\epsilon^4) )d s_2 d s_2 \nonumber\\
   && - \frac{1}{2}e^{-2(\psi+\phi)}y_1(2e^{2\phi}+y_1^2-(2\epsilon^4-\epsilon^2)y_2^2)d s_2
   dz_2+\nonumber\\
   &&
   \frac{1}{2}e^{-2(\psi+\phi)}y_2(2e^{2\phi}\epsilon^2+(2\epsilon^2+1)y_1^2+y_2^2\epsilon^2)d s_2
   dz_1 \nonumber\\
   &&
   + e^{-2(\psi+\phi)}(y_2^2 s_1\epsilon^2+2y_2  s_2
   y_1\epsilon^4+ s_1 y_1^2)d s_2 d\tilde t\nonumber\\
   && +
   \frac{1}{8}e^{-2(\psi+\phi)}(4e^{4\phi}+y_1^4+(2\epsilon^2+4)y_1^2y_2^2
   +y_2^4\epsilon^4+4e^{2\phi}(y_1^2+y_2^2))d s_1 d s_1\nonumber\\
   && - \frac{1}{2}e^{-2(\psi+\phi)}y_2(2e^{2\phi}+(2+\epsilon^2)y_1^2+y_2^2\epsilon^4)d s_1dz_2
   \nonumber\\
   &&- \frac{1}{2}e^{-2(\psi+\phi)}y_1(2e^{2\phi}+y_1^2-(2-\epsilon^2)y_2^2)d s_1 dz_1
   \nonumber\\
   &&-e^{-2(\psi+\phi)}(y_1^2  s_2 \epsilon^2- 2y_1 s_1 y_2 +  s_2 y_2^2\epsilon^4 )
   ds_1d\tilde t+
   \frac{1}{2}e^{-2(\psi+\phi)}(e^{2\phi}+y_1^2+y_2^2\epsilon^4)dz_2dz_2\nonumber\\
 && -e^{-2(\psi+\phi)}(2y_1 s_1 - 2  s_2 y_2\epsilon^4) dz_2
   d\tilde t+
   \frac{1}{2}e^{-4\psi}(e^{2\psi}+2z_2^2
   +e^{2(\psi-\phi)}(y_1^2+y_2^2))dz_1 dz_1\nonumber\\
   && +(2e^{-4\psi}( s_2^2\epsilon^2+ s_1^2)z_2 +2e^{-2(\psi+\phi)}(y_1 s_2\epsilon^2+ s_1
   y_2))dz_1d\tilde t\nonumber\\
   &&(e^{-4\phi}+2e^{-2(\phi+\psi)}(s_1^2+s_2^2\epsilon^4)+e^{-4\psi}(s_1^4+2s_1^2s_2^2\epsilon^2+s_2^4\epsilon^4))
    d\tilde t d\tilde t- 2 e^{-4\phi}y_1 d\tilde tdy_2\nonumber\\
   && + \frac{1}{2}e^{-4\phi}(e^{2\phi}+2y_1^2)dy_2 dy_2 +
   \frac{1}{2}e^{-2\phi}dy_1dy_1 + e^{-2(\phi+\psi)}y_1y_2(\epsilon^2-1)dz_1dz_2)\nonumber\\&&+
   \frac12
   e^{-2\psi}y_1y_2(1-\epsilon^2)(2+e^{-2\phi}(y_1^2+y_2^2\epsilon^2))ds_1ds_2,
   \end{eqnarray}
which can be compared with (\ref{bigmetrice=1}) for $\epsilon=1$.
For $\epsilon=0$ it becomes
\begin{eqnarray}
  ds^2&=&(d\phi^2+d\psi^2+\frac12 e^{-4\psi}(e^{-2(\phi-\psi)}+2s_2^2)d\tilde
  s_2^2\nonumber\\&&-
 \frac12 e^{-2(\phi+\psi)}y_1^2d\tilde s_2ds_1 +
 (e^{-2(\phi+\psi)}y_1+2e^{-4\psi}s_2z_2)d\tilde s_2
 dz_1\nonumber\\&&+
\frac18
 e^{-2(\phi+\psi)}(4e^{4\phi}+y_1^4+4y_1^2y_2^2+4e^{2\phi}(y_1^2+y_2^2))ds_1^2\nonumber\\&&
 -
 \frac12
 e^{-2(\phi+\psi)}y_1(2e^{2\phi}+y_1^2-2y_2^2)ds_1dz_1\nonumber
\\
  &&+
 \frac12
 e^{-4\psi}(e^{2\psi}+e^{-2(\phi-\psi)}(y_1^2+y_2^2)+2
 z_2^2)dz_1^2+
 \frac12 e^{-2\phi}dy_1^2)\nonumber\\
 \nonumber\\&&
 +e^{-4\psi}dt^2+
 e^{-4(\phi+\psi)}(e^{2\psi}+e^{2\phi}s_1^2)^2d \tilde t^2+
 \frac12e^{-4\psi}(e^{-2(\phi-\psi)}+2s_1^2)d\tilde s_1^2\nonumber\\&&+
 \frac18 e^{-2(\phi+\psi)}(2e^{2\phi}+y_1^2)^2 ds_2^2+
 \frac12 e^{-4\phi}(e^{2\phi}+2y_1^2)dy_2^2\nonumber\\&&+
 \frac12 e^{-2(\phi+\psi)}(e^{2\phi}+y_1^2)dz_2^2+
 2e^{-4\psi}s_1^2 dt d\tilde t+
 2e^{-4\psi}s_1 dt d\tilde s_1\nonumber\\&&  +
 2e^{-4\psi}s_2 dt d\tilde s_2 +
 2e^{-4\psi} z_2 dt dz_1+
 2e^{-2(\phi+2\psi)}s_1(e^{2\psi}+e^{2\phi}s_1^2)d\tilde t d\tilde
 s_1\nonumber\\&& +
 2e^{-4\psi}s_1^2 s_2 d\tilde t d\tilde s_2+
 2e^{-2(\phi+\psi)}s_1y_1y_2d\tilde t ds_1+
 e^{-2(\phi+\psi)}s_1y_1^2 d\tilde t ds_2\nonumber\\&&-
 2e^{-4\phi}y_1 d\tilde t dy_2+
 2e^{-2(\phi+2\psi)}s_1(e^{2\psi}y_2+e^{2\phi}s_1z_2)d\tilde t
 dz_1\nonumber\\&&-
 2e^{-2(\phi+\psi)}s_1y_1d\tilde t dz_2+
 2e^{-4\psi}s_1s_2 d\tilde s_1 d\tilde s_2+
 e^{-2(\phi+\psi)}y_1y_2 d\tilde s_1 ds_1\nonumber\\&&+
 \frac12 e^{-2(\phi+\psi)}y_1^2 d\tilde s_1 ds_2+
 e^{-4\psi}(e^{-2(\phi-\psi)}y_2+2 s_1 z_2)d\tilde s_1 dz_1-
 e^{-2(\phi+\psi)}y_1 d\tilde s_1 dz_2\nonumber\\&&+
 \frac12 e^{-2(\phi+\psi)}y_1(2e^{2\phi}+y_1^2)y_2ds_1 ds_2\nonumber\\&&-
 e^{-2(\phi+\psi)}y_2(e^{2\phi}+y_1^2)ds_1 dz_2-
 \frac12 e^{-2(\phi+\psi)}y_1(2e^{2\phi}+y_1^2)ds_2 dz_2\nonumber\\&&+
 \frac12 e^{-2(\phi+\psi)}y_1^2y_2 ds_2 dz_1-
 e^{-2(\phi+\psi)}y_1y_2dz_1dz_2\label{5lines}
\end{eqnarray}
We can compare the first 5 lines of (\ref{5lines}) with
(\ref{metric3}). They are different, but the  extra terms are zero
when imposing the constraints $$z_{2}=y_{2}=t=\tilde t=\tilde
s_1=s_2=0.$$ This means that there is an isometric embedding of
$\rSO(2,3)/(\rSO(2)\times \rSO(3))$ in the manifold with the
metric (\ref{5lines}). We can improve this result by making  use
of a generalized contraction, that gives  a simpler contracted
metric. We will do that in the next section.

\subsection{Generalized contractions: some examples.\label{gencon}}

\paragraph{Generalized contraction of $\rU(2,3)/(\rU(2)\times
U(3))$\\}

We consider the following decomposition of  $s_4$,
\begin{eqnarray*}&&\fs_4=\fg_0+\fg_1+\fg_2+\fg_3, \quad \hbox{where}\quad \\
&&\fg_0=\rspan\{H_+,H_-,S^{(1,1)}_2,S^{(1,-1)}_1,Y^1,Z^1,T^{(2,0)}\}\\
&&\fg_1=\rspan\{S^{(1,1)}_1\} \qquad \fg_2=\rspan\{T^{(0,2)},Z^2\}
\qquad \fg_3=\rspan\{S^{(1,-1)}_2,Y^2\}
\end{eqnarray*}and the linear map
\begin{equation*}
\begin{CD}\fs_4 @>\phi_\e>> \fs_4\\
e_0+e_1+e_2+e_3 @>>> e_0+\epsilon e_1+\epsilon^2 e_2+\epsilon^3
e_3\end{CD} \quad\hbox{with } e_i\in \fg_i.
\end{equation*}
Equation (\ref{contractedbracket}) gives a deformed bracket that
has a well defined limit when $\e\rightarrow 0$. We write here the
contracted bracket. The only surviving commutators from
(\ref{su23}) when $\e\rightarrow 0$  are
\begin{eqnarray*}
 &&[Z^{1},Y^{1}]=S^{(1,1)}_2\\
 &&[Y^{1},S^{(1,-1)}_1]=Z^{1}\\
  &&[H_+,Z^{i}]=Z^{i}\\
&& [H_-,Y^{i}]=Y^{i}, \end{eqnarray*} so the contracted algebra
has as a subalgebra
$$\fsolv\left(\frac {\rSO(2,3)}{\rSO(2)\times
\rSO(3)}\right)=\rspan\{H_+,H_-\}+\rspan\{Z^1,
Y^1,S^{(1,1)}_2,S^{(1,-1)}_1\}$$ in semidirect product with
$\R^6=\rspan\{Z^2,Y^2, T^{(0,2)},
T^{2,0},,S^{(1,1)}_1,S^{(1,-1)}_2 \}$, where the only generators
that act on $\R^6$ are $H_+$ and $H_-$.

We use the adjoint representation of the deformed algebra to
compute the deformed metric. The result is
 \begin{eqnarray}
   &&ds^2=
   d\phi^2 +
   d\psi^2+
   e^{-4\psi}dtdt+
   2e^{-4\psi} s_1 \epsilon dtd\tilde s_1 +
   2 e^{-4\psi} s_2 \epsilon^3 dtd\tilde s_2\nonumber\\
   && +2e^{-4\psi}z_2 \epsilon^2 dtdz_1 +
   2e^{-4\psi}(s_2^2\epsilon^2  + s_1^2 \epsilon^8)dtd\tilde t+
   \frac{1}{2}(e^{-2(\psi+\phi)}+ 2e^{-4\psi}s_1^2 \epsilon^2)d\tilde s_1 d\tilde s_1
   \nonumber\\
   && +2 e^{-4\psi} s_2  s_1 \epsilon^4 d\tilde s_1d\tilde s_2  +
   \frac{1}{2} e^{-2(\psi+\phi)}(y_1^2\epsilon^2+y_2^2\epsilon^8)d\tilde s_1d s_2+
   e^{-2(\psi+\phi)}y_1y_2 \epsilon^2 d\tilde s_1d s_1 \nonumber\\
   && -e^{-2(\psi+\phi)}y_1 \epsilon d\tilde s_1 dz_2+
   (2e^{-4\psi} s_1 z_2 \epsilon^3+e^{-2(\psi +\phi)}y_2\epsilon^2 ) d\tilde
   s_1dz_1\nonumber\\
   && + 2 e^{-4\psi} s_1 (e^{2(\psi-\phi)}\epsilon+ s_2^2\epsilon^9 + s_1^2\epsilon^3 ) d\tilde
   s_1d\tilde t +
 \frac{1}{2}(e^{-2(\psi+\phi)}+2e^{-4\psi} s_2^2\epsilon^6)d\tilde s_2 d\tilde
   s_2\nonumber\\
   &&+e^{-2(\psi+\phi)}y_1y_2 \epsilon^6 d\tilde s_2 d s_2
   -\frac{1}{2}e^{-2(\psi+\phi)}(y_1^2 +y_2^2\epsilon^6)d\tilde s_2 d s_1\nonumber\\
   &&+e^{-2(\psi+\phi)}y_2 \epsilon^5d\tilde s_2dz_2+
   (2e^{-4\psi}s_2 z_2\epsilon^5+e^{-2(\psi+\phi)}y_1)d\tilde s_2dz_1 \nonumber\\
   && + 2e^{-4\psi}s_2( e^{2(\psi-\phi)}\epsilon^5+ s_2^2\epsilon^{11} +  s_1^2\epsilon^5 )d\tilde s_2
   d\tilde t\nonumber\\
   && +
   \frac{1}{8}e^{-2(\psi+\phi)}(4e^{4\phi}+y_1^4\epsilon^4+(4\epsilon^{12}+2\epsilon^{10})y_1^2y_2^2+y_2^4\epsilon^{16}+4e^{2\phi}
   (y_1^2\epsilon^2+y_2^2\epsilon^{12}) )d s_2 d s_2 \nonumber\\
   && - \frac{1}{2}e^{-2(\psi+\phi)}y_1(2e^{2\phi}\epsilon+y_1^2\epsilon^3-(2\epsilon^{11}-\epsilon^9)y_2^2)d s_2
   dz_2+\nonumber\\
   &&
   \frac{1}{2}e^{-2(\psi+\phi)}y_2(2e^{2\phi}\epsilon^6+(\epsilon^2+2\epsilon^6)y_1^2+y_2^2\epsilon^{10})d s_2
   dz_1\nonumber\\
   && + e^{-2(\psi+\phi)}(y_2^2 s_1\epsilon^9+2y_2  s_2
   y_1\epsilon^{11}+ s_1 y_1^2\epsilon^3)d s_2 d\tilde t\nonumber\\
   && + \frac{1}{8}e^{-2(\psi+\phi)}(4e^{4\phi}+y_1^4+(4\epsilon^4+2\epsilon^6)y_1^2y_2^2
   +y_2^4\epsilon^{12}+4e^{2\phi}(y_1^2+y_2^2\epsilon^2))d s_1 d s_1\nonumber\\
   && - \frac{1}{2}e^{-2(\psi+\phi)}y_2(2e^{2\phi}\epsilon+(2\epsilon^3+\epsilon^5)y_1^2+y_2^2\epsilon^{11})d s_1dz_2\nonumber\\
   &&
   - \frac{1}{2}e^{-2(\psi+\phi)}y_1(2e^{2\phi}+y_1^2-(2\epsilon^4-\epsilon^6)y_2^2)d s_1 dz_1
   \nonumber\\
   &&-e^{-2(\psi+\phi)}(y_1^2  s_2\epsilon^5 - 2y_1 s_1 y_2 \epsilon^3+  s_2 y_2^2 \epsilon^{11})
   ds_1d\tilde t+
   \frac{1}{2}e^{-2(\psi+\phi)}(e^{2\phi}+y_1^2\epsilon^2+y_2^2\epsilon^{10})dz_2dz_2\nonumber\\
 && -e^{-2(\psi+\phi)}(2y_1 s_1\epsilon^2 - 2  s_2 y_2\epsilon^{10}) dz_2
   d\tilde t+
   \frac{1}{2}e^{-4\psi}(e^{2\psi}+2z_2^2\epsilon^4
   +e^{2(\psi-\phi)}(y_1^2+y_2^2\epsilon^2))dz_1 dz_1\nonumber\\
   && +(2e^{-4\psi}( s_2^2\epsilon^{10}+ s_1^2\epsilon^4)z_2 +2e^{-2(\psi+\phi)}(y_1 s_2\epsilon^3+ s_1
   y_2\epsilon^5))dz_1d\tilde t\nonumber\\
   &&+ (e^{4\psi}+e^{4\phi}(s_1^4\epsilon^4+2s_1^2s_2^2\epsilon^{10}+s_2^4\epsilon^{16})+2e^{2(\phi+\psi)}(s_1^2\epsilon^2+s_2^2\epsilon^{10})) d\tilde t d\tilde t
   - 2 e^{-4\phi}y_1 \epsilon d\tilde t dy_2\nonumber\\
   && + \frac{1}{2}e^{-4\phi}(e^{2\phi}+2y_1^2\epsilon^2)dy_2 dy_2 +
   \frac{1}{2}e^{-2\phi}dy_1dy_1 +
   e^{-2(\phi+\psi)}y_1y_2(\epsilon^5-\epsilon^3)dz_1dz_2\nonumber\\&&
   +\frac12e^{-2\psi}y_1y_2(2(\epsilon^2-\epsilon^6)+(\epsilon^4-\epsilon^6)(y_1^2+y_2^2\epsilon^6))ds_1ds_2.
   \end{eqnarray}

And for $\epsilon=0$ it becomes
\begin{eqnarray*}
ds^2&=& (d\phi^2 + d\psi^2 +\frac{1}{2}e^{-2(\psi+\phi )}d\tilde
s_2^2  -\frac{1}{2}e^{-2(\psi+\phi)}y_1^2 d\tilde s_2 d
s_1\\
&& +e^{-2(\psi+\phi)}y_{1}d\tilde s_2dz_{1}
 + \frac{1}{8}e^{-2(\psi+\phi)}(4e^{4\phi}+y_1^4+4e^{2\phi}y_1^2)d
 s_1^2
 \\&&- \frac{1}{2}e^{-2(\psi+\phi)}y_1(2e^{2\phi}+y_{1}^2
)d s_1 dz_{1} +\frac{1}{2}e^{-4\psi}(e^{2\psi}
+e^{2(\psi-\phi)}y_{1}^2)dz_{1}^2 +
\frac{1}{2}e^{-2\phi}dy_{1}^2)\\\\&&+ e^{-4\psi}dt^2+
e^{-4\phi}d\tilde t^2 +\frac12 e^{-2(\phi+\psi)}d\tilde s_1^2 +
\frac12 e^{2(\phi-\psi)}ds_2^2\\&& + \frac12 e^{-2\phi} dy_2^2
+\frac12 e^{-2\psi}dz_2^2
\end{eqnarray*}
The first three lines reproduce (\ref{metric3}) for $n=1$, and the
rest of the terms are flat up to factors  $ e^{(a\phi+b\psi)}$.
The physical meaning of this limit remains unclear at this moment,
but it relates two different sigma models in what can be a
generalized procedure of integrating out some modes.

\paragraph{Generalized contraction of $\rU(2,1+n)/(\rU(2)\times
\rU(1+n))$\\} We show here another example of generalized
contraction that has an application in a physically interesting
theory.

Let us denote $H_1=H_++H_-$. Then the commutation rules of $\fs_2$
are
\begin{eqnarray*}[H_1,Z^{1a}]=Z^{1a}, \quad
[H_1,Y^{1a}]=Y^{1a},\quad [H_1,S_2^{(1,1)}]=2S_2^{(1,1)}, \quad
[Z^{1a}, Y^{1b}]=S_2^{(1,1)}. \label{s2}\end{eqnarray*} Consider
the subalgebras of $\fs_4$
$$\fs_2'=\rspan\{H_+, Z^{ia},  T^{(2,0)}\}\qquad \fs_2''=\rspan\{H_-, Y^{ia},
T^{(0,2)}\},$$ with commutation rules
\begin{eqnarray*}&&[H_+,Z^{ia}]=Z^{ia}, \quad
[H_+,T^{(2,0)}]=2T^{(2,0)},\quad
[Z^{ia},Z^{jb}]=\delta^{ab}\epsilon^{ij}T^{(2,0)},\quad
\mathrm{for}\; \;\fs_2',\\&& [H_-,Y^{ia}]=Y^{ia}, \quad
[H_-,T^{(0,2)}]=2T^{(0,2)},\quad
[Y^{ia},Y^{jb}]=\delta^{ab}\epsilon^{ij}T^{(0,2)},\quad
\mathrm{for} \; \fs_2''.
\end{eqnarray*}
We have that $\fs_2\simeq \fs_2'\simeq\fs_2''$ but $[\fs_2',
\fs_2'']\neq 0$, so $\fs_2'\oplus \fs_2''$ is not a subalgebra of
$\fs_4$. Nevertheless,
 one  can find a generalized contraction of $\fs_4$ which has $\fs_2'\oplus
 \fs_2''$ as a subalgebra.  We consider the decomposition,
 \begin{eqnarray*}
 &\fs_4=\fg_0+\fg_{1}+\fg_{2}\\\\
&\fg_0=\rspan\{H_+,H_-\},\quad  \fg_{1}=\rspan \{Y^{ia}, Z^{ia},
S_\alpha^{(1,1)} \},\quad \fg_{2}=\{T^{(0,2)},
T^{(2,0)},S_\alpha^{(1,-1)}\},
\end{eqnarray*} and the linear map
\begin{equation}
\begin{CD}\fs_4@>\phi_\epsilon>> \fs_4\\
e_0+e_1+e_2@>>>e_0+\epsilon e_1+ {\epsilon^2} e_2\end{CD}\qquad
e_i\in \fg_i. \label{contractionmap}
\end{equation}
 The contracted Lie algebra, with commutator given by
 \ref{contractedbracket})
 is well
defined. It is worthy to see the commutators of the contracted
algebra:
\begin{eqnarray}
&&[Z^{ia},Z^{jb}]_\e=\epsilon^{ij}\delta^{ab}T^{(2,0)}\nonumber\\
&&[Y^{ia},Y^{jb}]_\e=\epsilon^{ij}\delta^{ab}T^{(0,2)}\nonumber\\&&
[Z^{ia},Y^{jb}]_\e=\epsilon \;\delta^{ab}(\delta^{ij}S_2^{(1,1)}+\epsilon^{ij}S_1^{(1,1)})\nonumber\\
&&[Y^{ia},S_1^{(1,-1)}]_\e=\epsilon^2 \; Z^{ia}\nonumber\\
&&[Y^{ia},S_2^{(1,-1)}]_\e= \epsilon^2\; \epsilon^{ij}Z^{ja}\nonumber\\
&&[T^{(0,2)}, S_\alpha^{(1,-1)}]_\e=\epsilon^3\; 2S_\alpha^{(1,1)}\nonumber\\
&&[S_\alpha^{(1,1)},S_\beta^{(1,-1)}]_\e=\epsilon \; \delta_{\alpha\beta}T^{(2,0)}\nonumber\\
&&[H_+, Z^{ia}]_\e=Z^{ia}\nonumber\\
&&[H_-, Y^{ia}]_\e=Y^{ia}\label{su23deformed}
\end{eqnarray}
showing explicitly $\fs_2'\oplus \fs_2''$ as a subalgebra when
$\e\rightarrow 0$. We use the adjoint representation to compute
the metric, as in the previous examples.
The result is

\begin{eqnarray}
   &&ds^2=
   d\phi^2 +
   d\psi^2+
   e^{-4\psi}dtdt+
   2e^{-4\psi}\epsilon s_1 dtd\tilde s_1 +
   2 e^{-4\psi}\epsilon s_2 dtd\tilde s_2\nonumber\\
   && +2e^{-4\psi}z_{2a} dtdz_{1a} +
   2e^{-4\psi}\epsilon^4(s_2^2 + s_1^2)dtd\tilde t+
   \frac{1}{2}(e^{-2(\psi+\phi)}+ 2e^{-4\psi}\epsilon^2 s_1^2)d\tilde s_1 d\tilde s_1
   \nonumber\\
   && +2 e^{-4\psi}\epsilon^2 s_2  s_1 d\tilde s_1d\tilde s_2  +
   \frac{1}{2} e^{-2(\psi+\phi)}\epsilon^3(y_{1}^2+y_{2}^2)d\tilde s_1d s_2+
   e^{-2(\psi+\phi)}\epsilon^3y_{1a}y_{2a} d\tilde s_1d s_1 \nonumber\\
   && -e^{-2(\psi+\phi)}\epsilon y_{1a} d\tilde s_1 dz_{2a}+
   \epsilon(2e^{-4\psi} s_1 z_{2a} +e^{-2(\psi +\phi)}y_{2a} ) d\tilde
   s_1dz_{1a}\nonumber\\
   && + 2 e^{-4\psi}\epsilon^3 s_1 (e^{2(\psi-\phi)}+ \epsilon^2(s_2^2 + s_1^2) ) d\tilde
   s_1d\tilde t +
 \frac{1}{2}(e^{-2(\psi+\phi)}+2e^{-4\psi}\epsilon^2 s_2^2)d\tilde s_2 d\tilde
   s_2\nonumber\\
   &&+e^{-2(\psi+\phi)}\epsilon^3y_{1a}y_{2a} d\tilde s_2 d s_2
   -\frac{1}{2}e^{-2(\psi+\phi)}\epsilon^3(y_{1}^2 +y_{2}^2)d\tilde s_2 d s_1\nonumber\\
   &&+e^{-2(\psi+\phi)}\epsilon y_{2a}d\tilde s_2dz_{2a}+
   \epsilon(2e^{-4\psi}s_2 z_{2a}+e^{-2(\psi+\phi)}y_{1a})d\tilde s_2dz_{1a} \nonumber\\
   && + 2e^{-4\psi}\epsilon^3 s_2( e^{2(\psi-\phi)}+\epsilon^2( s_2^2 +  s_1^2) )d\tilde s_2
   d\tilde t\nonumber\\
   && +
   \frac{1}{8}e^{-2(\psi+\phi)}(4e^{4\phi}+ 4e^{2\phi}
   \epsilon^4(y_{1}^2+y_{2}^2)+ 4\epsilon^6(y_{1a}y_{2a})^2+\epsilon^6(y_{1}^2+y_{2}^2)
   (y_{1b}^2+y_{2b}^2) )d s_\alpha d s_\alpha \nonumber\\
   && - \frac{1}{2}e^{-2(\psi+\phi)}\epsilon^2(2e^{2\phi}y_{1b}+\epsilon^2
   (-2(y_{1a}y_{2a})y_{2b}+(y_{1}^2
   +y_{2}^2)y_{1b}))d s_2
   dz_{2b}\nonumber\\&&+
   \frac{1}{2}e^{-2(\psi+\phi)}\epsilon^2(2e^{2\phi}y_{2b}+\epsilon^2(2(y_{1a}y_{2a})y_{1b}+
   (y_{1}^2+y_{2}^2)y_{2b}))d s_2
   dz_{1b}\nonumber\\
   && + e^{-2(\psi+\phi)}\epsilon^6(y_{2}^2 s_1+2y_{2a}  s_2
   y_{1a}+ s_1 y_{1}^2)d s_2 d\tilde t\nonumber\\
    && - \frac{1}{2}e^{-2(\psi+\phi)}\epsilon^2(2e^{2\phi}y_{2b}+\epsilon^2(2(y_{1a}y_{2a})y_{1b}+(y_{1}^2+y_{2}^2)y_{2b}))
    d s_1dz_{2b}\nonumber \\&&
   - \frac{1}{2}e^{-2(\psi+\phi)}\epsilon^2(2e^{2\phi}y_{1b}+\epsilon^2(-2(y_{1a}y_{2a})y_{2b}+
   (y_{1a}^2+y_{2a}^2)y_{1b}))d s_1 dz_{1b}
   \nonumber\\
   &&-e^{-2(\psi+\phi)}\epsilon^6(y_{1a}^2  s_2 - 2y_{1a} s_1 y_{2a} +  s_2 y_{2a}^2 )
   ds_1d\tilde t\nonumber\\&&-
\frac{1}{2}e^{-2(\phi+\psi)}\epsilon_{ij}\epsilon_{mn}\epsilon^2(y_{ia}y_{jb})dz_{ma}dz_{nb}\nonumber\\&&
   +\frac{1}{2}e^{-2(\psi+\phi)}(e^{2\phi}\delta_{ab}+\epsilon^2(y_{1a}y_{1b}+y_{2a}y_{2b}))dz_{2a}dz_{2b}\nonumber\\
 && -e^{-2(\psi+\phi)}\epsilon^4(2y_1 s_1 - 2  s_2 y_2) dz_2
   d\tilde t\nonumber\\&&+
   \frac{1}{2}e^{-4\psi}(e^{2\psi}\delta_{ab}+2z_{2a}z_{2b}+e^{2(\psi-\phi)}
   \epsilon^2(y_{1a}y_{1b}+y_{2a}y_{2b}))dz_{1a} dz_{1b}\nonumber\\
   && +\epsilon^4(2e^{-4\psi}( s_2^2+ s_1^2)z_{2a} +2e^{-2(\psi+\phi)}(y_{1a} s_2+ s_1
   y_{2a}))dz_{1a}d\tilde t\nonumber\\
   &&+ e^{-4(\psi+\phi)}(e^{4\psi}+e^{2(\phi+\psi)}\epsilon^6( s_1^2 +
   s_2^2)+e^{4\phi}\epsilon^8( s_1^2 +
   s_2^2)^2
   ) d\tilde t d\tilde t- 2 e^{-4\phi}y_{1a} d\tilde tdy_{2a}\nonumber\\
   && + \frac{1}{2}e^{-4\phi}(e^{2\phi}\delta_{ab}+2y_{1a}y_{1b})dy_{2a} dy_{2b} +
   \frac{1}{2}e^{-2\phi}dy_{1a}dy_{1a} \label{bigmetric}
   \end{eqnarray}
In the contraction limit $\epsilon \to 0$ the metric reduces to

\begin{eqnarray*}ds^2=&&\bigl(d\phi^2+e^{-4\phi}
d\tilde t d\tilde t- 2 e^{-4\phi}y_{1a} d\tilde tdy_{2a} +\\&&
\frac{1}{2}e^{-4\phi}(e^{2\phi}\delta_{ab}+2y_{1a}y_{1b})dy_{2a}
dy_{2b} +
   \frac{1}{2}e^{-2\phi}dy_{1a}dy_{1a}\bigr)+\\&&
\bigl(d\psi^2+e^{-4\psi}dtdt+2e^{-4\psi}z_{2a}dtdz_{1a}+\\&&\frac
12 e^{-4\psi}(e^{2\psi}\delta_{ab}+
2z_{2a}z_{2b})dz_{1a}dz_{1b}+\frac 12
e^{-2\psi} dz_{2a}dz_{2a}\bigr)+\\
 &&+\frac
12e^{-2(\psi+\phi)}d\tilde s_\alpha d\tilde s_\alpha+\frac 12
e^{-2(\psi+\phi)}ds_\alpha ds_\alpha.
\end{eqnarray*} By  comparison with (\ref{metric2}) we can see, after a suitable renaming of the coordinates and a rescaling by a global constant factor,
that this is the metric  on
$$\left(\frac{\rSU(1,1+n)}{\rU(1+n)} \times
\frac{\rSU(1,1+n)}{\rU(1+n)}\right)\ltimes \R^4. $$



 \section{Super Higgs mechanism in Supergravity: geometric interpretation}
 We consider
 an $N=2$ supergravity model coupled to $n+2$ hypermultiplets and $n+1$ vector
 multiplets. This model comes from a compactification of type IIB supergravity on
 the $N=4$ orientifold  $T^6/\Z_2$ \cite{fp1,kst}. Such model, when certain fluxes
 are turned on, has an $N=3$ phase
 obtained after the integration of the massive gravitino multiplet. The theory
 describing the $N=3$ massles modes
can be further Higgsed to an $N=2$ phase by turning on other
suitable fluxes and further integration \cite{adfl2}. The scalar manifold for the
$N=2$ theory is \cite{fp}
$$\cM_Q\times \cM_{SK}=\frac{\rU(2,2+n)}{\rU(2)\times \rU(2+n)}
\times \frac {\rU(1,1+n)}{\rU(1)\times \rU(1+n)}.$$ Here $n$
refers to the brane degrees of freedom. The special geometry and
symplectic basis that describe this model have been discussed in
Ref. \cite{dflv}.

The $N=2$ model can also be Higgsed to N=1,0 phases by still turning
on fluxes. This corresponds in the supergravity language to gauge
two translational isometries of the quaternionic manifold. In the
parametrization that we have used before (\ref{bigmetric}) it is
manifest that the coordinates  $\tilde s_\alpha$
 correspond to
two translational isometries, generated by $S_\alpha^{(1,1)} $, and
we used the two bulk vector fields to gauge them.

When the gauge interactions are switched on, the $\sigma$-model
lagrangian gets modified by the
 minimal coupling prescription, with
$$d\tilde s_\alpha \to D \tilde s_\alpha = d\tilde s_\alpha + k_{\alpha ,
\Lambda}A^\Lambda ,\qquad \Lambda=0,1,\dots n+1$$ and we may
choose the constants $k_{1,0}\neq 0$,  $k_{2,1}\neq 0$ and the
rest zero. Then the Higgs mechanism takes place, as we described
in Section \ref{integration}, with the $\tilde s_\alpha$
contributing to the longitudinal components of the massive vectors
$$\hat A^0_\mu= A_\mu^0+\frac 1{k_{1,0}}\partial_\mu\tilde s_1,\qquad  \hat A^1_\mu=
A_\mu^1+\frac 1{k_{2,1}}\partial_\mu\tilde s_2.$$ From Eq.
(\ref{vielbein}) we can see that the kinetic term of these modes
is
\begin{eqnarray*}
ds^2= &&\cdots +e^{-4\psi}\left(s_\alpha D\tilde s_\alpha  +
(s_1^2+s_2^2)
   d\tilde t +dt +z_{2a}
   dz_{1a}\right)^2+\\&&
\frac 1 2 e^{-2(\psi +\phi)}\left( D\tilde s_1
+y_{1a}y_{2a}ds_1+\frac 12(y_1^2+y_2^2)ds_2 + 2s_1d\tilde
t-\epsilon_{ij}y_{ia}dz_{ja}\right)^2+\\&& \frac 12e^{-2(\psi
+\phi)}\left( D\tilde s_2 +y_{1a}y_{2a}ds_2-\frac
12(y_1^2+y_2^2)ds_1 +2s_2d\tilde
t+\delta_{ij}y_{ia}dz_{ja}\right)^2+\cdots
   \end{eqnarray*}
After the substitution
 $$D\tilde s\alpha\rightarrow k_{\alpha,\Lambda}\hat A^\Lambda=B^\Lambda$$
 the kinetic term for the vectors remains unchanged, while mass
 terms appear for the vectors $\hat A^\Lambda$, with masses
 $k_{0,1}$ and
 $k_{1,2}$. The modes $\tilde s_\alpha$ disappear from the
 Lagrangian.

 In the large mass limit, the
massive fields $B_\mu^\Lambda= m \hat A_\mu^\Lambda$ appear in the
Lagrangian through expressions of the type
$$(B_\mu+f_\mu)^2,$$
where $f_\mu$ is some interaction of the massless modes \cite{fp}.
The $B$'s are Lagrange multipliers and their equations of motions
make these terms to vanish.

The $N=2$ gauged theory has a scalar potential stabilizing two
additional scalars which in our parametrization, correspond  to the
coordinates $s_\alpha$ \cite{adfl1}. These fields acquire also a
mass through the potential. In the large mass limit, these fields
become Lagrange multipliers and the potential is such that their
field equations set them to zero.

After performing these integrations, the metric becomes the one of
the symmetric space
$$\frac{\rU(1,1+n)}{\rU(1)\times \rU(1+n)}\times \frac{\rU(1,1+n)}{\rU(1)\times \rU(1+n)}.$$

Now we see that this example fits with the contraction performed in
Section \ref{gencon}. We can see that the terms set to zero in the
metric by taking the limit $\e\rightarrow 0$ are precisely the terms
eliminated by the integration procedure.  The modes that have become
massive are the modes in $\R^4$ in the decomposition
$$\fsolv\left(\frac{\rU(2,2+n)}{\rU(2)\times \rU(2+n)}\right)\rightarrow
\fsolv\left(\frac{\rU(1,1+n)}{\rU(1)\times \rU(1+n)}\times
\frac{\rU(1,1+n)}{\rU(1)\times \rU(1+n)}\right)\ltimes\R^4.$$
Since $\R^4$ is an invariant subgroup of the contracted group, the
quotient
\begin{eqnarray*}\fsolv\left(\frac{\rU(1,1+n)}{\rU(1)\times
\rU(1+n)}\times \frac{\rU(1,1+n)}{\rU(1)\times
\rU(1+n)}\right)\ltimes\R^4/\R^4\\\approx\fsolv\left(\frac{\rU(1,1+n)}{\rU(1)\times
\rU(1+n)}\times \frac{\rU(1,1+n)}{\rU(1)\times
\rU(1+n)}\right),\end{eqnarray*} is a (solvable) group, associated
to the symmetric space. So in the geometrical picture  the
integration of the massive modes is again  modeled by a
contraction and a quotient by an invariant subgroup.

\appendix


\section{\label{translational}About solvable Lie algebras and
translational isometries} A Lie algebra $\fs$ is solvable if the
chain of ideals
$$\fs^{(0)}=\fs, \quad \fs^{(1)}=[\fs, \fs], \dots ,
\fs^{(p)}=[\fs^{(p-1)}, \fs^{(p-1)}], \dots$$ has $\fs^{(p+1)}=0$
for some integer $p$. It is possible to prove \cite{va} that a Lie
algebra $\fs$ is solvable if and only if there is a chain of
ideals $\fid_{i+1}\subset\fid_{i}$ with $\fid_{i}/\fid_{i+1}$ an
abelian algebra, $\fid_0=\fs$ and $\fid_{p+1}=0$ for some $p$. It
is clear that $\fid_{p}=0$ is an abelian ideal.

\begin{example}\end{example}As an example, let us consider
$\fs=\fs_4$ so
\begin{eqnarray*}\fid_0&=&\fs_4\\
\fid_1=[\fid_0,\fid_0]&=&\rspan\{ Z^{ia}, Y^{ia}, T^{2,0},
T^{0,2},
S_\alpha^{(1,1)},S_\alpha^{(1,-1)}\} \\
\fid_2=[\fid_1,\fid_1]&=&\rspan\{ Z^{ia},  T^{2,0}, T^{0,2},
S_\alpha^{(1,1)}\}\\
\fid_3=[\fid_2,\fid_2]&=&\rspan\{ T^{2,0}\}.\end{eqnarray*} Notice
that it is possible  to  substitute $\fid_3$ in the chain by the
maximal abelian ideal
$$\fid'_3=\rspan\{ T^{2,0}, T^{0,2}, S_\alpha^{(1,1)},Z^{1a} \},$$
or by this other one (with the same dimension)
$${\fid''}_3=\rspan\{ T^{2,0}, T^{0,2}, S_\alpha^{(1,1)},Z^{2a} \},$$
so the chain is not unique.

\hfill$\blacksquare$

\bigskip

 One can also show that $\fs^{(1)}$ is a nilpotent Lie
algebra. The unique simply connected group associated to a
nilpotent Lie algebra is exponential (the exp map is a
diffeomorphism of the Lie algebra into the Lie group) \cite{va}.

\bigskip

Let $\fg$ be a Lie algebra and $\ft$ an abelian subalgebra. Let
$\{X_i\}$ be a basis  of $\ft$ and $\{Y_\alpha\}$ a basis of a
complementary space to $\ft$. In a neighborhood of the identity,
we have the exponential map \begin{equation}L(u^i,
v^\alpha)=e^{u^iX_i}e^{v^\alpha
Y_\alpha}.\label{parametrization}\end{equation} The Maurer-Cartan
form is
$$L^{-1}dL= e^{-v^\alpha Y_\alpha}X_i e^{v^\alpha Y_\alpha}du^i +
 e^{-v^\alpha Y_\alpha}d(e^{v^\alpha Y_\alpha}).$$
 From this expression, one can see that the local expression of the
 Maurer-Cartan form does not depend on the coordinates $u^i$.
 Whenever the group $G$ with Lie algebra $\fg$ is diffeomorphic to
 $\R^n\times M$, with $\R^n$ parametrized  by $u^i$ (in other words,  the coordinates $u^i$ are
 global), we will say that the generators $X_i$ are {\it translational isometries}.
 \bigskip

 We consider now the solvable algebras associated to the non compact
 symmetric
 spaces by the Iwasawa decomposition and explore the translational isometries in the corresponding
 symmetric spaces. The solvable Lie algebras are always a semidirect
 product
 $$\fs=\fa\ltimes\fn,$$
 where $\fa$ is abelian (it contains the non compact Cartan
 elements) and
 $\fs^{(1)}=\fn$ is the nilpotent part. The non compact symmetric spaces are simply connected,
 so they are, in each case, the unique simply
 connected group associated to the corresponding solvable algebra.
 We denote it by
 $$S=A\ltimes N,\quad \hbox{with}\quad \mathrm{Lie}(A)=\fa\;\;
 \mathrm{and}\;\;
 \mathrm{Lie}(N)=\fn,$$ $A$ and $N$ being simply connected  as well\cite{he}
 (and hence, exponential). As a manifold,
 \begin{equation}S=A\times N= \exp(\fa)\times \exp(\fn).\label{factors}\end{equation}
 \bigskip

 Let us now consider the factor $N$ in (\ref{factors}). We want to
 prove that the generators in the abelian ideal are translational
 isometries. Let $\fn=\fn_1+\fn_2$, with $\fn_1$ an abelian ideal and $\fn_2$ any complementary subspace.
 We have that the map
 \begin{eqnarray*}\begin{CD}\fn@>>> N\\
 (X_1,X_2)@>>>\exp(X_1+X_2)\end{CD}\end{eqnarray*}
 is a diffeomorphism. We want to show that, equally, the map
\begin{eqnarray*}\begin{CD}\fn@>>> N\\
 (X_1,X_2)@>>>\exp X_1\exp X_2\end{CD}\end{eqnarray*} is a diffeomorphism.
 It is enough to prove that obvious that any element $\exp(Y_1+Y_2)$ can
 be written as $\exp X_1\exp X_2$  for some $X_i\in
 \fn_i$. We notice that
 $$\exp X_1\exp X_2=\exp (X_1+X_2 +\frac 1 2[X_1,X_2]+\cdots)=\exp
 (X_2 +X'_1), \;\hbox{with}\; X'_1\in \fn_1.$$
 We take $Y_2=X_2$, and the equation $Y_1=X'_1$ can be solved for
 some $X_1\in \fn_1$.

 \section*{Acknowledgements}

M. A. Ll. wants to thank the Physics and Mathematics Departments
at UCLA and the Department of Physics, Theory Division, at CERN
for their hospitality during the realization of this work.

S. F. wants to thank the Departament de F\'{\i}sica Te\`orica of
the Universitat de Val\`encia  for its kind hospitality during the
realization of this work.

 The work of S.F. has been supported in
part by the D.O.E. grant DE-FG03-91ER40662, Task C, and in part by
the European Community's Human Potential Program under contract
HPRN-CT-2000-00131 Quantum Space-Time, in association with INFN
Frascati National Laboratories.

The work of M. A. Ll. and O. M.  has been supported by the
research grant BFM 2002-03681 from the Ministerio de Ciencia y
Tecnolog\'{\i}a (Spain) and from EU FEDER funds.

The work of M. A. Ll. has also been supported by D.O.E. grant
DE-FG03-91ER40662, Task C.

M. A. Ll. and O. M. want to thank V. S. Varadarajan and J. A. de
Azc\'arraga for helpful discussions .

 \end{document}